\newcommand{\ergs}{\,erg\,s$^{-1}$}
\newcommand{\kms}{km\,s$^{-1}$}

\newcommand{\Msol}{M$_{\odot}$}
\newcommand{\Rsol}{R$_{\odot}$}
\newcommand{\Teff}{$T_{\rm eff}$\,}

\newcommand{\logg}{$\log g$\,}

\newcommand{\Lbol}{\rm L$_{\rm bol}$}
\newcommand{\Mbol}{\rm M$_{\rm bol}$}
\newcommand{\Lx}{$\rm L_{\rm X}$}

\newcommand{\figheight}{6.0cm}
\documentclass[mathleft,a4paper]{an}
\usepackage{graphicx}
\usepackage{times}
\usepackage[authoryear]{natbib}
\begin{document}

% The following seven commands are intended for editorial usage and should be ignored by
% the author(s).
%\Pagespan{000}{}% Document's page range. 
% If second parameter is left empty, the last page is computed automatically.
\Yearpublication{2010}%
\Yearsubmission{2010}%
\Month{}%   
\Volume{}%  
\Issue{}% 
% \DOI{This.is/not.aDOI}% 

\title{The supergiant optical counterpart of ULX P13 in NGC7793\thanks{Based on optical observations obtained at ESO under programme 084.D.0881(A)}}

\author{C. Motch \inst{1}\fnmsep\thanks{Corresponding author:
  \email{christian.motch@unistra.fr}\newline}
\and  M.W. Pakull\inst{1}
\and  F. Gris\'e\inst{2}
\and  R. Soria\inst{3}
}
\titlerunning{The supergiant optical counterpart of ULX P13 in NGC7793}
\authorrunning{C. Motch et al.}
\institute{
CNRS, Universit\'e de Strasbourg, Observatoire Astronomique, 11 rue de l'Universit\'e, F-67000 Strasbourg, France
\and 
Department of Physics and Astronomy, University of Iowa, Van Allen Hall, Iowa City, IA 52242, USA
\and 
Mullard Space Science Laboratory, University College London, Holmbury St Mary, Surrey RH5 6NT, UK
}

\received{ }
\accepted{ }
%\publonline{later}


\abstract{%
We have identified the optical counterpart of the ULX source P13 in the nearby spiral galaxy NGC 7793. The object is a V $\sim$ 20.5 mag star, ten times brighter than any other established counterpart of a ULX in nearby galaxies. Medium resolution optical spectroscopy carried out in 2008 and 2009 with the ESO-VLT reveals the presence of narrow high order Balmer, HeI and MgII absorption lines indicating a late B type supergiant companion star with mass between 10 and 20\,\Msol. Stellar H$\beta$ and HeII\,$\lambda$4686 emission lines are also seen superposed on the photospheric spectrum. We detect different patterns of radial velocity variations from the emission and absorption lines over a time interval of one month. The velocity of the high order Balmer absorption lines changes by $\sim$ 100\,\kms\ while the H$\beta$ and HeII\,$\lambda$4686 emission components vary by about the same amount but with a different phasing. Assuming that the observed velocity changes trace the motion of the mass-donor star and of the X-ray source implies a mass of the accreting black hole in the range of 3 to 100\,\Msol\ with a most probable value of $\sim$ 10 to 20\,\Msol. We expect an orbital period in the range of 20 to 40 days based on the low density of the supergiant star. P13 is likely in a short-lived, and thus rare high X-ray luminosity evolutionary state associated with the ascension of the donor star onto the supergiant stage.}

\maketitle

\section{The ULX P13 in NGC 7793}

Many galaxies harbour non-nuclear X-ray sources with \Lx\
higher than $10^{39}$ and up to several $10^{40}$\ergs, generally known as ultraluminous X-ray sources (ULXs). X-ray variability shows that the majority are accreting compact
objects. If the Eddington limit is not violated, or if no strong radiation beaming occurs, such
isotropic luminosities imply compact stellar masses up to a few 100\,\Msol. However, presently available stellar evolutionary models at solar me\-tallicity do not
predict compact remnants that massive. Even intermediate-mass black holes with 
M $\sim$ $10^3$\,\Msol\ have be\-en suggested, based on X-ray
spectral and timing argu\-ments \citep[see e.g.][]{miller2004}.
Determining black hole masses is therefore a key issue for understanding the X-ray emission mechanism and the evolutionary status of ULXs. 

NGC 7793 is an Sd galaxy member of the Sculptor Gr\-oup of galaxies located at a distance of 3.9\,Mpc \citep{karachentsev2003}. 
{\it ROSAT} PSPC observations \citep{read1999} have revealed that its X-ray emission is dominated by a single source, P13, located at the southern edge of the galaxy and exhibiting evidence of variability over a 5 month time interval. {\it Chandra} revisited NGC 7793 in 2003 for 49\,ks \citep{pannuti2006}, providing an accurate source position at RA = 23:57:51.01, Dec =  $-$32:37:26.62 (J2000) with a formal 90\% confidence radius of 0.4\arcsec. A simple po\-wer law does not fit well the {\it Chandra} ACIS-S spectrum ($\chi^{2}$\,=\,201 for 181\,dof). However, the addition of a soft or of a hot blackbody decreases the $\chi^{2}$ to acceptable values ($\chi^{2}$\,$\sim$\,185 for 179\,dof in both cases). The corresponding two best fits are $\Gamma$ = 1.10 $\pm$ 0.06, kT = 0.18 $\pm$ 0.03\,keV and $\Gamma$ = 2.14 $\pm$ 0.28, kT = 1.72 $\pm$ 0.13\,keV. With \Lx $\sim$ 4 $\times$10$^{39}$\-\ergs, P13 is a bona fide ULX. A blue V$\sim$20.5 stellar object falls right in the middle of the small {\it Chandra} error circle, 
leaving no doubt about the identification of P13 counterpart, the brightest ULX optical counterpart presently known (Pakull et al. 2010, in prep).

\begin{figure*}
\centerline{\includegraphics[height=14cm,width=5.45cm,angle=-90,clip=true,bb= 40 80 560 790]{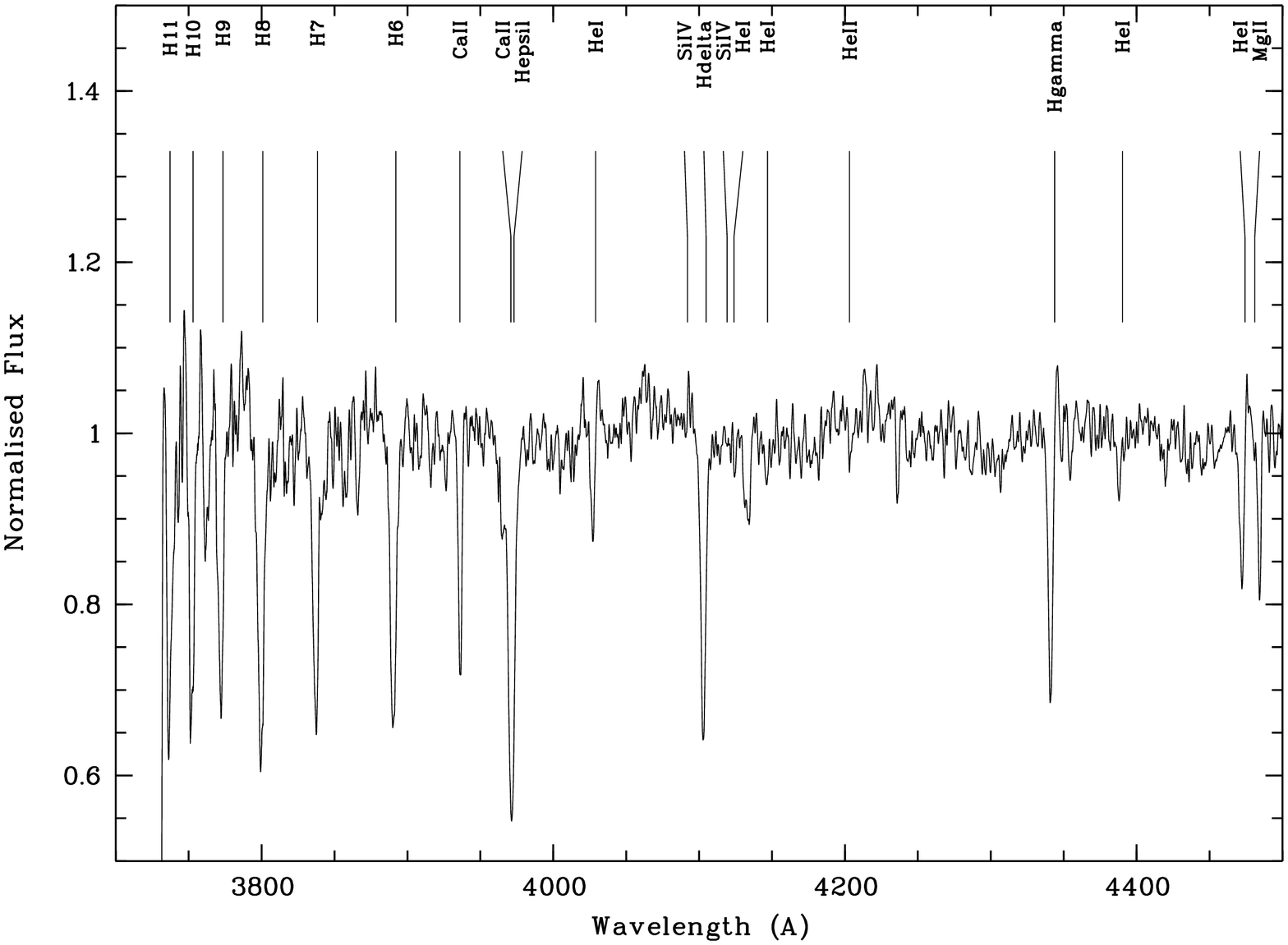}}
\centerline{\includegraphics[height=14cm,width=5.45cm,angle=-90,clip=true,bb= 40 80 560 790]{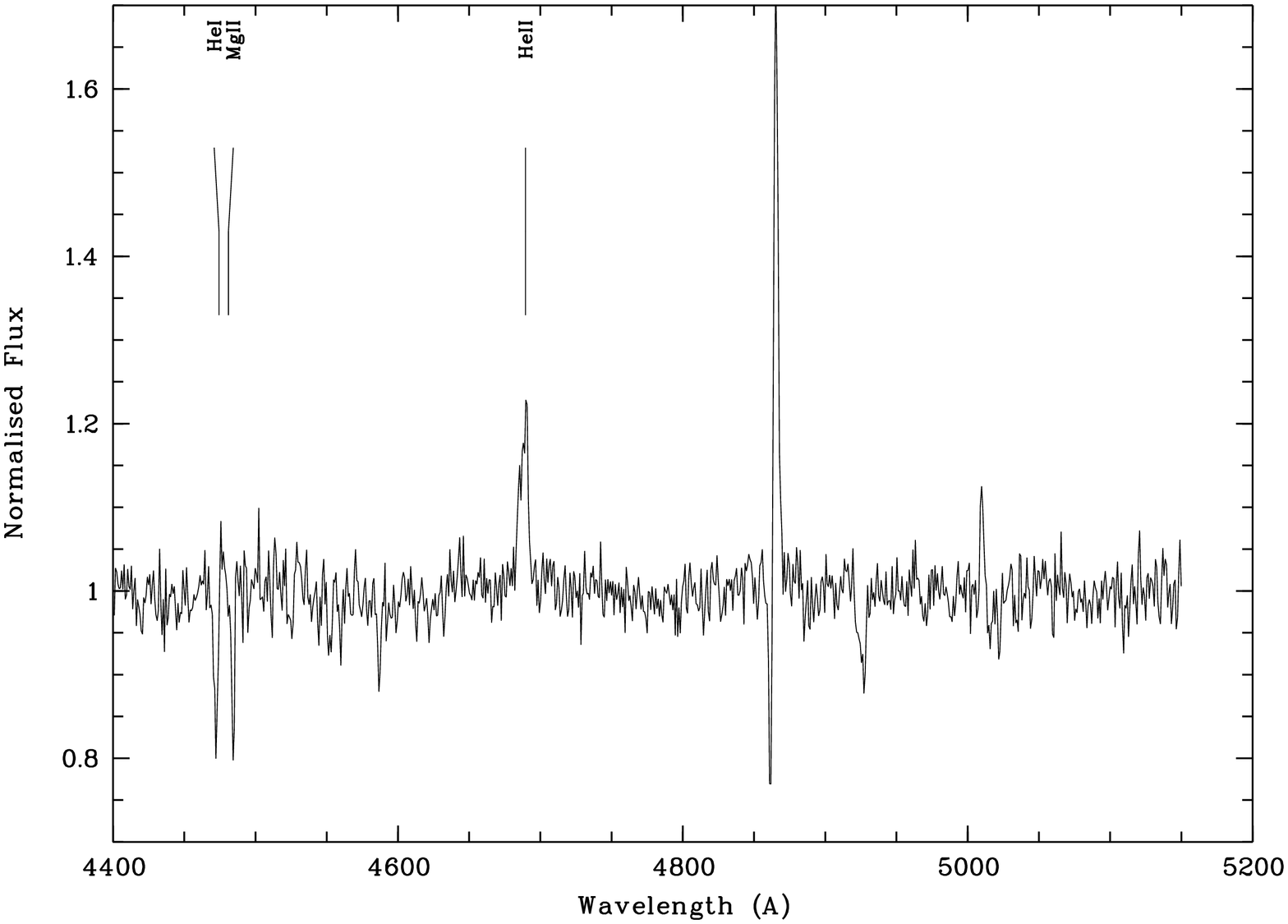}}
\caption{Optical spectrum of P13 obtained by summing all 2009 observations (total exposure time of 3h). The continuum has been normalized to unity. Photospheric lines include the high order Balmer absorption lines, some HeI lines and the MgII line. The HeII line is clearly seen in emission. Balmer emission is strong at H$\beta$ and added to the photospheric absorption line yields a marked P Cyg profile.}
\label{mean_spec}
\end{figure*}

\section{Optical properties}

In 2008 and 2009 we obtained ESO-VLT observations with the aim of deriving the spectral type of the companion star and of measuring the dynamical masses of the binary components. Unfortunately, very few of the scheduled observations were eventually executed in 2008 and 2009. Additional runs are expected to be carried out in 2010. On 2008 November 23 and 27, two 40\,min long spectra were obtained with FORS1 and the 600B Grism providing a spectral resolution of $\sim$ 600 over the wavelength range 3500-6000\AA. The five 2009 40\,min long exposures were obtained with a higher resolution of $\sim$ 1600 over the 3700-5200\AA\ range using FORS2 and grism 1200B. Observations took place on November 17 and December 9, 11, 13 and 14. The rectified sum of all optical spectra obtained in 2009 shown in Fig. \ref{mean_spec} reveals many deep unresolved Balmer and HeI absorption lines. Strong emission from the HeII $\lambda$4686 and H$\beta$ lines is superposed onto the photospheric absorption spectrum. The ratio of the HeI\,$\lambda$4471 to MgII $\lambda$4481 lines indicates a B8 spectral type \citep[\Teff\ $\sim$ 14,000K;][]{walborn1990} while the high optical brightness of the star Mv $\sim$ $-$7 implies a supergiant luminosity class. Our 2008 spectra display a strong Balmer jump indicating \logg\ $\sim$ 2 in agreement with the luminosity class. Masses and radii of such stars are not accurately known. Evolutionary tracks with rotation from \cite{meynet2000} predict masses of the order of 20\,\Msol\ for our observed \Mbol\ of $ \sim$ $-8$. However, masses as low as 10\,\Msol\ have also been reported for such stars \cite[see e.g.][]{markova2008}. The high optical luminosity implies a stellar radius of 40 to 60\,\Rsol. Note that in spite of its very large optical luminosity, the total emission of P13 remains dominated by its high energy output. The X-ray to stellar bolometric luminosity ratio \Lx/\Lbol $\sim$\,6 suggests that some X-ray heating effects might be observable. However, the impact of X-ray heating on the derived photospheric \Teff\ is difficult to assess without knowing the geometry of the system.

\section{Radial velocities}

We applied the standard ESO {\em gasgano} + {\em fors-kit-4.1} pipe\-line to the 2009 FORS2 data to create wavelength calibrated spectra corrected for bias and flat-field. Remaining spectral distortions in the near UV region were removed using additional arc lines at short wavelengths. The final wavelength scale is accurate to 2\- \kms \ while the small variation of the target position with\-in the slit from one observation to the other yielded  a scatter of less than 5\,\kms. 

Fig. \ref{mean_spec} shows high order Balmer lines mostly in absorption, whereas emission components become conspicuous at H$\gamma$ and below. Using a cross-correlation me\-thod we monitored the velocity of the 5 Balmer lines contained in the 3750 to 3900\AA\ interval. We tested several different templates: a) the straight average of all spectra with and without velocity correction, b) a B8I spectrum extracted from the \cite{walborn1990} atlas and c) a TLUTSY B\-STAR06 model \citep{lanz2007} with \Teff\,=\,15,000 K and \logg\,=\,2.0, degraded to the FORS2 resolution. Results are not sensitive to the actual template used or to the inclusion of additional absorption lines at slightly longer wavelengths such as Ca H\&K, H$\epsilon$ and H$\delta$. Errors on velocities were derived from sub-exposures measurements and estimated to be of the order of 7\,\kms. 

In order to monitor the behaviour of Balmer emission components, we subtracted from the observed spectra a template photospheric spectrum shifted at the measured velocity of the high order Balmer absorption lines. We tested again two of the templates considered for the cross-correla\-tion method, namely the \cite{walborn1990} atlas spectrum and the TLUTSY model without finding noticeable effects on the results. The subtracted spectra only display emissions at H$\gamma$ and H$\beta$. The centroid of the H$\beta$ emission line was measured by fitting a Gaussian profile. Unfortunately, the H$\gamma$ emission is too weak to provide useful velocity information. The HeII $\lambda$4686 line has a broad resolved profile on the top of which a narrower component appears on occasion. Since the line does not overlap with any strong photospheric feature, we directly measured its position on the original spectra by fitting Gaussian profiles to the broad or narrow components or using a cross-correlation method. Again, all approaches gave quite similar results.  

\begin{figure}
\includegraphics[width=\figheight,height=\columnwidth,clip=true,angle=-90,bb= 40 50 560 600]{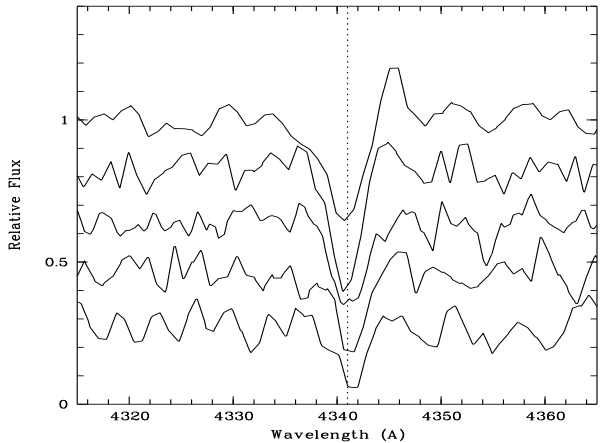}
\caption{Variation of the H$\gamma$ absorption line profile with time. Spectra have been shifted in flux for clarity. 
From top to bottom, Nov. 17, Dec. 9, 11, 13 and 14.}
\label{hgamma}
\end{figure}

We find that the velocity of the absorption features chan\-ges by 105.5 $\pm$ 13\,\kms \ over the 27 day time interval between the first and last observation. The velocity shift is also visible in the H$\gamma$ line plotted in Fig. \ref{hgamma}, although this particular line is clearly contaminated by residual emission. 

\begin{figure}
\includegraphics[width=\figheight,height=\columnwidth,clip=true,angle=-90,bb= 40 50 560 600]{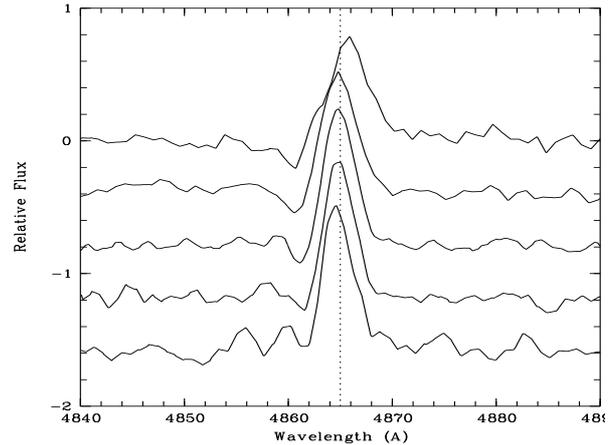}
\caption{Variation of the H$\beta$ emission line profile with time. A TLUTSY/BSTAR06 model spectrum shifted at the velocity of the high order Balmer absorption lines was subtracted from the observed spectrum. Spectra are shifted in intensity for clarity. From top to bottom, Nov. 17, Dec. 9, 11, 13 and 14.}
\label{hbeta}
\end{figure}

\begin{figure}
\includegraphics[width=\figheight,height=\columnwidth,clip=true,angle=-90,bb= 40 50 560 600]{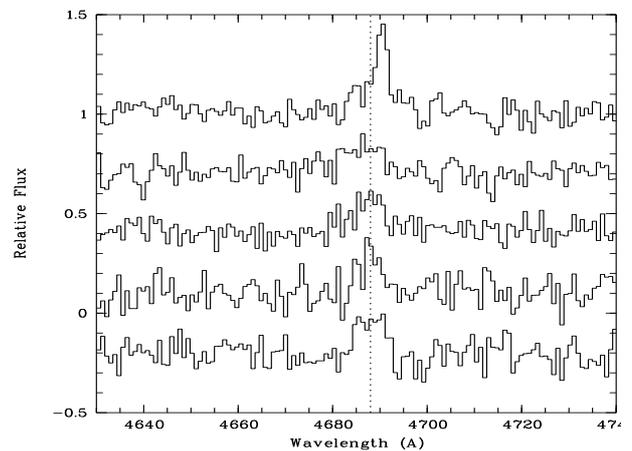}
\caption{Variation of the HeII $\lambda$4686 emission line profile with time. Spectra are shifted in intensity for clarity. From top to bottom, Nov. 17, Dec. 9, 11, 13 and 14.}
\label{heII}
\end{figure}

Most of the velocity change of the H$\beta$ emission line occurs between the first and second data points separated by 22 days (see Fig. \ref{hbeta}) with a total amplitude of 73.4$\pm$11.0 \kms. A P\,Cyg profile is patent in several spectra. However, the line equivalent width (EW) remains stable at a value of $\sim$ $-$3.5\AA.  In contrast, the HeII $\lambda$4686 line exhibits larger profile and velocity variations (see Fig. \ref{heII}). Ignoring the second observation in which the line was particularly broad and weak, the total velocity amplitude of the HeII line is 104$\pm$15\,\kms, i.e. comparable to that of the Balmer absorption lines. The line EW varies from $-1$ to $-2.4$ \AA. 

\begin{figure}
\includegraphics[width=\columnwidth,height=5.7cm,clip=true,angle=0,bb= 0 0 535 405]{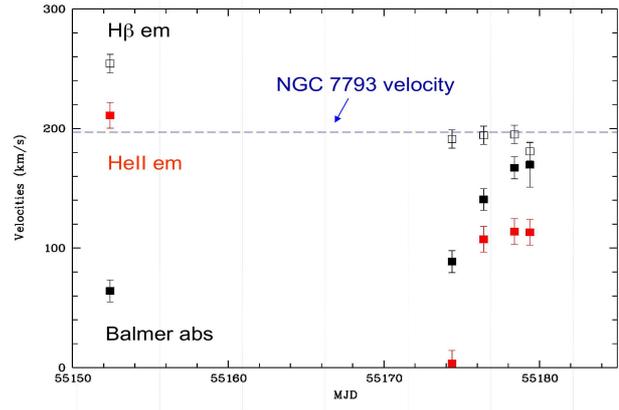}
\caption{Radial velocity curves of the high order Balmer absorption lines (black filled squares) of the H$\beta$ emission (open squares) and of the He $\lambda4686$ emission line (red filled squares). The horizontal line shows the velocity of the nearby interstellar gas.}
\label{RV}
\end{figure}

The radial velocity curves plotted in Fig. \ref{RV} show a rather complex pattern which is hard to interpret considering the scarcity of the available data. One can remark, however, that between the single November 17 observation and the mean of the December observations, the velocities of the emission and absorption components roughly varied in opposite directions. Such a behaviour would be expected in a double line spectroscopic binary. The velocity of the absorption lines probably reflects to a large extent the orbital motion of the mass-donor star. Unfortunately, its smooth variation during the 4 day long run in December is not accompanied by any correlated change in the H$\beta$ or He II line velocities. This hints at an orbital phasing different from 180\degr\ between absorption and emission lines, or worse, at a mostly random variation. P13 might thus be not as clean as one would have liked to measure dynamical masses of both components. It is fairly possible that none of the emission lines accurately follows the motion of the compact object, neither in amplitude nor in phase. Many supergiant stars exhibit Balmer emission lines with P\,Cyg profiles produced in massive winds. In the case of P13, the difference in velocity between the peak emission and absorption is $\sim$ 250-330 \kms, fully consistent with the escape velocity from a 20\,\Msol star having a 60\,\Rsol\ radius. Therefore, emission from an X-ray photoionized stellar wind could well add to the light originating from the accretion disc and considerably alter the H$\beta$ radial velocity curve. 

Similar considerations apply to the HeII line. Its velocity profile might include emissions from the X-ray heated companion hemisphere or from various structures of the accretion disc. 
In fact, previous optical spectroscopic observations of ULXs cast some doubt on the reliability of the HeII $\lambda$4686 line as a tracer of the motion of the accreting component \citep{roberts2010}. Most ULXs are likely to undergo noticeable X-ray heating effects with an irradiated to unirradiated optical flux ratio closer to that seen in galactic low-mass X-ray binaries than shown by classical high-mass X-ray binaries (HMXRBs) \citep[][]{copperwheat2005,grise2008}. This could imply, for instance, that the HeII $\lambda$4686 line does not follow the velocity of the accreting component but rather that of the disc bulge at the point where the stream of matter escaping L1 impacts the edge of the disc \citep[see e.g.][]{pearson2006}. Moreover, although for the first time in a ULX we detect photospheric absorption lines, in other ULXs spectral signatures from the companion stars have not yet been found \citep[][]{roberts2010}.  

However, in the case of P13, the roughly opposite velocity variations of the HeII + H$\beta$ and high order Balmer absorption lines suggest that these lines may still be used to constrain to some extent the dynamical parameters.
In addition, the moderately high \Lx/\Lbol\ of $\sim$ 6 is similar to that of the X-ray luminous HMXRB LMC X-4 in which the HeII $\lambda$4686 line does trace the motion of the neutron star \citep[K$_{\rm X}$ $\sim$ K$_{\rm HeII}$;][]{kelley1983}.

Adopting the Roche lobe overflow formalism of \cite{eggleton1983} yields orbital periods of about 20 to 40 days for a large range of mass ratios and for stellar radii between 40 and 60\,\Rsol. The fact that the total velocity amplitudes of Balmer absorption and HeII lines are both larger than $\sim$ 100\,\kms\ implies a mass ratio $q$ = M$_{\rm opt}$/M$_{\rm X}$ in the range of 0.2 to 3.2. Values of K$_{\rm X}$ $\sim$ K$_{\rm opt}$ $\sim$ 50\,\kms\ are reached for $i$ $\sim$ 30\degr. A mass ratio close to 1 is supported by the nearly equal amplitude of the absorption and emission lines radial velocity curves and therefore hints at a black hole mass of $\sim$ 10 - 20\,\Msol. Taken at face value, our observations do not support the presence of an intermediate mass black hole in P13.

\section{Discussion and Conclusions}

Determining whether ULXs harbour stellar mass or in some cases intermediate mass black holes can only be settled by measuring orbital periods and dynamical masses. However, the optical faintness of most ULXs is a challenge for me\-dium resolution spectroscopic monitoring, even using 8m-class telescopes. P13 is the first ULX in which optical spectral signatures of the mass-donor star are detected. These lines provide both a reliable indication of the spectral type and luminosity class of the companion star and at the same time means to measure accurate orbital velocities.   The high optical luminosity and the late B spectral type of the companion star clearly demonstrate that the star has left the main sequence and is now rapidly evolving towards a red supergiant. This evolutionary stage corresponds to the most X-ray luminous phases of any ULX \citep{rappaport2005} and should last no more than $\sim$ 10$^{6}$ years and may be as short as a few 10$^{4}$ years \citep{patruno2008}. In comparison, the life time of ULXs in which the mass-donor star is a sub-giant is much longer ($\sim$ 10$^{7}$ years). Population models indeed confirm that systems like P13 are rare and should account for only $\sim$ 5\% of all ULXs \citep{rappaport2005}. 
%Determining whether first Roche lobe contact happened during the supergiant phase of was already occurring during the main sequence phase, evolutionary case B or AB, will be possible once orbital parameters are known. 

%\acknowledgements

%\vskip -1cm

\vspace{-2pt}
\bibliography{p13}

\begin{thebibliography}{16}
\expandafter\ifx\csname natexlab\endcsname\relax\def\natexlab#1{#1}\fi

\bibitem[{{Copperwheat} {et~al.}(2005){Copperwheat}, {Cropper}, {Soria}, \&
  {Wu}}]{copperwheat2005}
{Copperwheat}, C., {Cropper}, M., {Soria}, R., \& {Wu}, K. 2005, \mnras, 362,
  79

\bibitem[{{Eggleton}(1983)}]{eggleton1983}
{Eggleton}, P.~P. 1983, \apj, 268, 368

\bibitem[{{Gris{\'e}} {et~al.}(2008){Gris{\'e}}, {Pakull}, {Soria}, {Motch},
  {Smith}, {Ryder}, \& {B{\"o}ttcher}}]{grise2008}
{Gris{\'e}}, F., {Pakull}, M.~W., {Soria}, R., {et~al.} 2008, \aap, 486, 151

\bibitem[{{Karachentsev} {et~al.}(2003){Karachentsev}, {Grebel}, {Sharina},
  {Dolphin}, {Geisler}, {Guhathakurta}, {Hodge}, {Karachentseva}, {Sarajedini},
  \& {Seitzer}}]{karachentsev2003}
{Karachentsev}, I.~D., {Grebel}, E.~K., {Sharina}, M.~E., {et~al.} 2003, \aap,
  404, 93

\bibitem[{{Kelley} {et~al.}(1983){Kelley}, {Jernigan}, {Levine}, {Petro}, \&
  {Rappaport}}]{kelley1983}
{Kelley}, R.~L., {Jernigan}, J.~G., {Levine}, A., {Petro}, L.~D., \&
  {Rappaport}, S. 1983, \apj, 264, 568

\bibitem[{{Lanz} \& {Hubeny}(2007)}]{lanz2007}
{Lanz}, T. \& {Hubeny}, I. 2007, \apjs, 169, 83

\bibitem[{{Markova} \& {Puls}(2008)}]{markova2008}
{Markova}, N. \& {Puls}, J. 2008, \aap, 478, 823

\bibitem[{{Meynet} \& {Maeder}(2000)}]{meynet2000}
{Meynet}, G. \& {Maeder}, A. 2000, \aap, 361, 101

\bibitem[{{Miller} {et~al.}(2004){Miller}, {Fabian}, \& {Miller}}]{miller2004}
{Miller}, J.~M., {Fabian}, A.~C., \& {Miller}, M.~C. 2004, \apjl, 614, L117

\bibitem[{{Pannuti} {et~al.}(2006){Pannuti}, {Schlegel}, \&
  {Lacey}}]{pannuti2006}
{Pannuti}, T.~G., {Schlegel}, E.~M., \& {Lacey}, C.~K. 2006, in IAU Symp., Vol.
  230, Populations of High Energy Sources in Galaxies, ed. {E.~J.~A.~Meurs \&
  G.~Fabbiano}, 197--198

\bibitem[{{Patruno} \& {Zampieri}(2008)}]{patruno2008}
{Patruno}, A. \& {Zampieri}, L. 2008, \mnras, 386, 543

\bibitem[{{Pearson} {et~al.}(2006){Pearson}, {Hynes}, {Steeghs}, {Jonker},
  {Haswell}, {King}, {O'Brien}, {Nelemans}, \& {M{\'e}ndez}}]{pearson2006}
{Pearson}, K.~J., {Hynes}, R.~I., {Steeghs}, D., {et~al.} 2006, \apj, 648, 1169

\bibitem[{{Rappaport} {et~al.}(2005){Rappaport}, {Podsiadlowski}, \&
  {Pfahl}}]{rappaport2005}
{Rappaport}, S.~A., {Podsiadlowski}, P., \& {Pfahl}, E. 2005, \mnras, 356, 401

\bibitem[{{Read} \& {Pietsch}(1999)}]{read1999}
{Read}, A.~M. \& {Pietsch}, W. 1999, \aap, 341, 8

\bibitem[{{Roberts} {et~al.}(2010){Roberts}, {Gladstone}, {Goulding},
  {Swinbank}, {Ward}, {Goad}, \& {Levan}}]{roberts2010}
{Roberts}, T., {Gladstone}, J., {Goulding}, A., {et~al.} 2010, \an, these
  proceedings

\bibitem[{{Walborn} \& {Fitzpatrick}(1990)}]{walborn1990}
{Walborn}, N.~R. \& {Fitzpatrick}, E.~L. 1990, \pasp, 102, 379

\end{thebibliography}

\end{document}